\documentclass[11pt]{JACoW-GSI}
\usepackage{graphicx,wrapfig,cite,fancyhdr,url,moreverb,lineno,rotating,lscape,amsmath,calc,multicol}

\newcommand{\psip}{\psi '}
\newcommand{\jpsi}{J/\psi}

\newcommand{\mumu}{\mu^{+}\mu^{-}}
\newcommand{\pipi}{\pi^{+}\pi^{-}}
\newcommand{\piz}{\pi^{0}}

\newcommand{\kpi}{$K\pi^-$}

\newcommand{\psipi}{$\psi\pi^-$}
\newcommand{\jpsipi}{$J/\psi\pi^-$}
\newcommand{\psitwospi}{$\psi^{\prime}\pi^-$}
\newcommand{\mpsitwospi}{$m_{\psi^{\prime}\pi^-}$}

\newcommand{\Ksone}{$K^{\ast}(892)$}
\newcommand{\Kstwo}{$K^{\ast}_2(1430)$}

\newcommand{\z}{$Z(4430)^-$}
\newcommand{\costhk}{$\cos\theta_K$}

\input{babarsym}

\begin{document}

\title{\begin{flushleft} 
\small{\textnormal{SLAC-PUB-14353}}\\
\small{\textnormal{SNUQ2C-11001}}\\ 
\end{flushleft}   \bf{ \boldmath {The {\it $Z$} Charmoniumlike Mesons}}}

\author[1]{Arafat Gabareen Mokhtar\thanks{mokhtar@slac.stanford.edu}}
\author[2]{Stephen Lars Olsen\thanks{solsen@hep1.snu.ac.kr}}
\affil[1]{SLAC National Accelerator Laboratory, Stanford, California
94309, USA} \affil[2]{Seoul National University, KOREA}
\vspace{18.0mm}

\maketitle

\vspace{3.0cm}

\centerline{\bf Abstract} A brief review of the experimental situation
concerning the electrically-charged charmoniumlike meson candidates,
$Z^-$, is presented.

\section{Introduction}

The Belle Collaboration reported peaks in the
$\psi^{\prime}\pi^-$~\cite{psiprime} and $\chi_{c1}\pi^-$ invariant
mass distributions~\cite{conj} in $B\to \psi^{\prime}\pi^-
K$~\cite{belle_z4430} and $B\to\chi_{c1}\pi^- K$~\cite{belle_z14050},
respectively. If these peaks are meson resonances, they would have a
minimal quark substructure of $c\bar{c}d\bar{u}$ and be unmistakeably
exotic. However, even though the Belle signals have more than
5$\sigma$ statistical significance, the experimental situation remains
uncertain in that none of these peaks have yet been confirmed by other
experiments. An analysis by the \babar\ Collaboration of
$B\to\psi^{\prime}\pi^- K$ neither confirms nor contradicts the Belle
claim for the $Z(4430)^-\to \psi^{\prime}\pi^-$~\cite{babar_z4430}. In
the \babar\ analysis, $B\to J/\psi\pi^- K$ decays were also studied,
and no evidence for $Z(4430)^-\to J/\psi\pi^-$ was found.

In this paper, we review and compare Belle and \babar\ results on
searches for charged charmoniumlike states. An abbreviated version of
this review is contained in Ref.~\cite{qwg_2010}.

\section{The Belle observation}
\begin{figure}[htb]
\centering \includegraphics*[width=80mm]{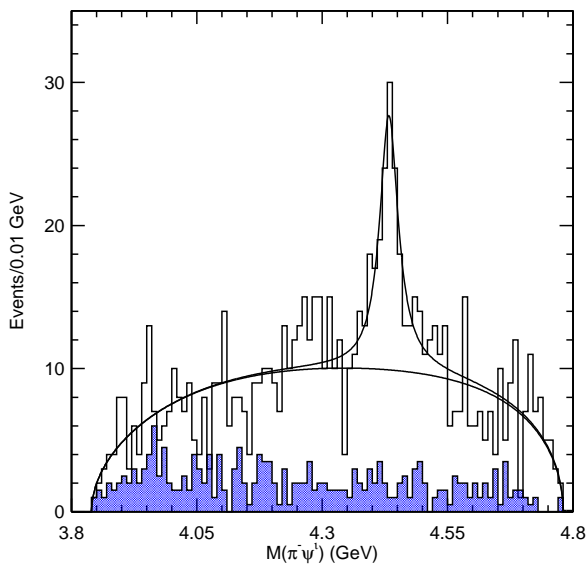}
\caption{From Belle~\cite{belle_z4430}, the \psitwospi\ invariant mass
distribution for $B\to\psi^{\prime}\pi^- K$ decays after applying a
veto on the \Ksone\ and \Kstwo\/. The open histogram shows the data,
and the shaded histogram represents the scaled results from the
background. The solid curves are the fit results.}
\label{fig:z4430_mpipsip}
\end{figure}

\begin{figure}[htb]
\centering
\includegraphics*[width=80mm]{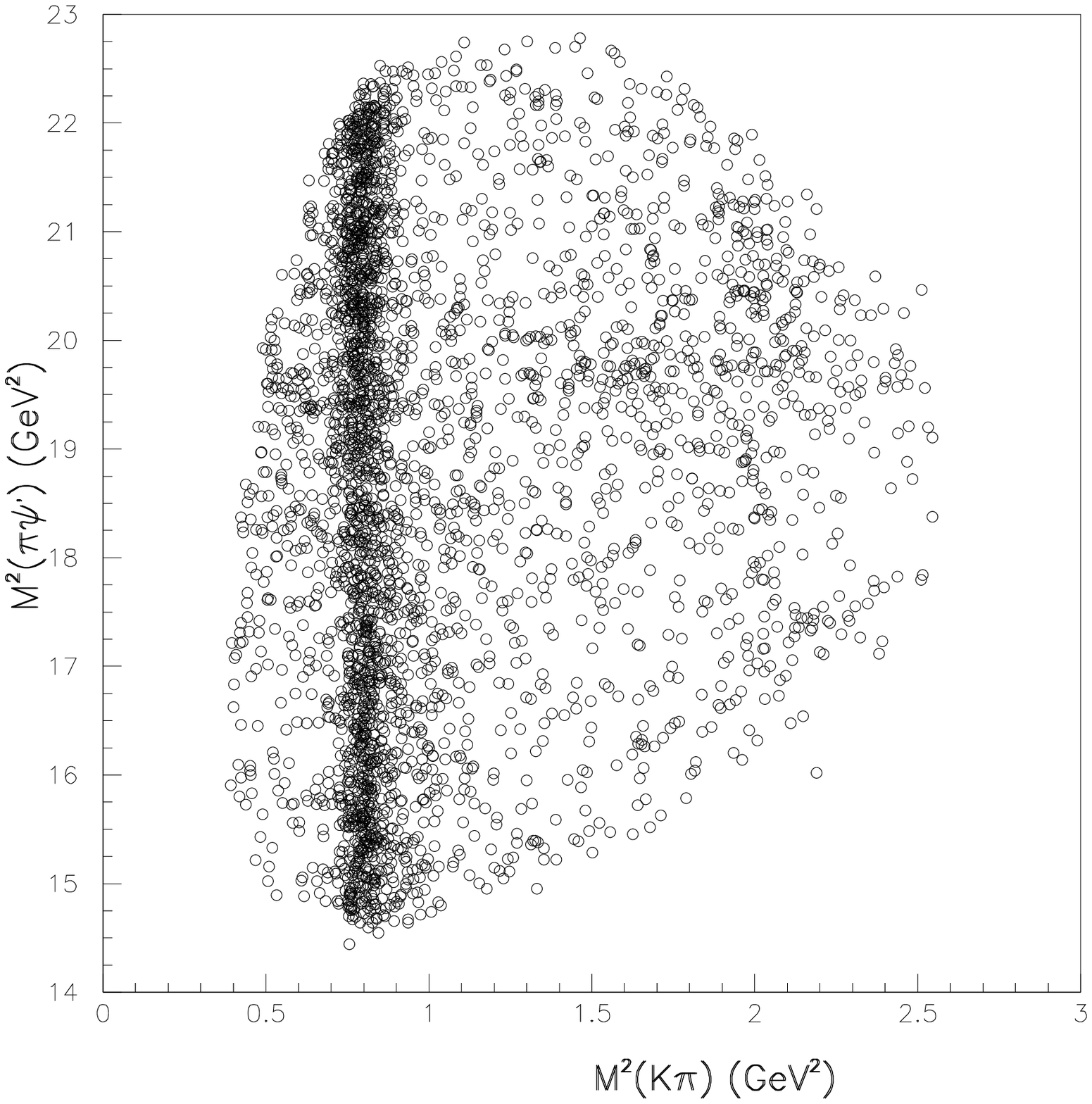}
\caption{From Belle~\cite{belle_z4430}, the $M^2_{K\pi^-}$ (horizontal)
{\it vs.} $M^2_{\psi^{\prime}\pi^-}$ (vertical) Dalitz plot
distribution for candidate $B\to\psi^{\prime}\pi^- K$ events.}
\label{fig:z4430_dalitz}
\end{figure}

Belle's original \z\ signal~\cite{belle_z4430} is the sharp peak in
the \psitwospi\ invariant mass distribution from $B\to
\psi^{\prime}\pi^- K$ decays shown in
Fig.~\ref{fig:z4430_mpipsip}. The Belle analysis is based on a data
sample that is equivalent to an integrated luminosity of 605
fb$^{-1}$. Figure~\ref{fig:z4430_dalitz} shows the Dalitz plot for
$B\to \psi^{\prime}\pi^- K$ candidates, where vertical bands for
\Ksone\/$\to$\kpi\ and \Kstwo\/$\to$\kpi\ are evident and the \z\
shows up as a horizontal band of events between
$M^2_{\psi^{\prime}\pi^-}=19$ and 20 GeV$^2$/c$^4$. (In the
$\psi^{\prime}\pi^-$ invariant mass distribution of
Fig.~\ref{fig:z4430_mpipsip}, the the $K^\ast$ bands are suppressed by
cuts on the \kpi\ masses~\cite{belle_z4430}).

A binned maximum-likelihood fit to the \psitwospi\ mass distribution
using a Breit Wigner (BW) resonance function for the signal and a
polynomial background gives a mass of $M=4433\pm 4\pm 2$ \mevcc\ and
total width of $\Gamma = 45^{+18~~+30}_{-13~~-13}$ \mev\/, with
estimated statistical significance of more than $6\sigma$. Consistent
signals are seen in various subsets of the data: {\it i.e.}  for both
the $\psi^{\prime}\to \ell^+\ell^-$ ($\ell=e$, or $\mu$) and
$\psi^{\prime}\to J/\psi\pi^+\pi^-$ ($J/\psi\to\ell^+\ell^-$)
subsamples~\cite{belle_z4430}.

\section{A reflection from $K\pi$ dynamics?}
Identifying resonant structures in the \psitwospi\ channel in
three-body $B\to \psi^{\prime}\pi^- K$ decays has to be done with care
because of the possibility that dynamics in the \kpi\ channel can
cause mass structures in the \psitwospi\ invariant mass distribution
that have no relation to \psitwospi\ dynamics. Energy-momentum
conservation imposes a tight correlation between the decay angle
($\theta_{\pi^-}$) in the \kpi\ system~\cite{theta-pi_def} and the
\psitwospi\ invariant mass, and this results in
$M^2_{\psi^{\prime}\pi^-}$ being very nearly proportional to
$\cos\theta_{\pi^-}$. Consequently, interference between different
partial waves in the \kpi\ system can produce peaks in 
$M_{\psi^{\prime}\pi^-}$ that are purely ``reflections'' of structures
in $\cos\theta_{\pi^-}$ and have nothing to do with \psitwospi\
dynamics. However, in the kinematically allowed \kpi\ mass range for
$B\to\psi^{\prime}\pi^-K$ decay, only $S$-, $P$- and $D$-waves
are significant, and this limited set of partial waves
can only produce fake \psitwospi\ mass peaks at a discrete set of mass
values.

\begin{figure}[htb]
\centering \includegraphics*[width=80mm]{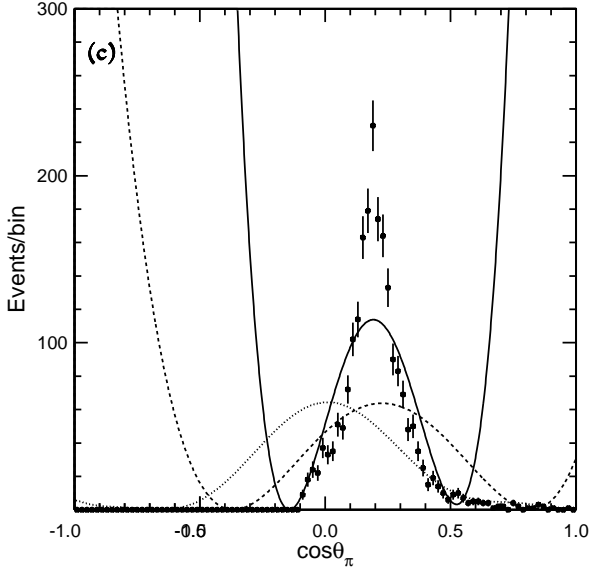}
\caption{From Ref.~\cite{solsen_charm09}, the histogram shows the
$\cos\theta_{\pi^-}$ distribution for a MC-generated \psitwospi\
resonance with $M=4.43$ \gevcc\ and $\Gamma = 0.05$ \gev\/. The curves
show the results of attempts to produce a peak in the vicinity of the
data with interfering $S$-, $P$- and $D$-waves in the \kpi\ channel.}
\label{fig:z4430_coskpi}
\end{figure}

In the case of the \z\/, the \psitwospi\ peak mass corresponds to
$\cos\theta_{\pi^-}\simeq 0.25$, and it is not possible to produce a
peak near this $\cos\theta_{\pi^-}$ value with any combination of
interfering $L=$0, 1, and 2 partial waves without introducing
additional, larger structures at other $\cos\theta_{\pi^-}$ values.
This is illustrated in Fig.~\ref{fig:z4430_coskpi}, where the
histogram shows the distribution of $\cos\theta_{\pi^-}$ values for a
MC sample of $B\to Z(4430)^-K$, \z\/$\to$\psitwospi\ events where the
\z\ mass and width closely correspond to Belle's reported
values~\cite{belle_z4430}. The curves in the figure show the results
of trying to make a peak at the same location with interfering $S$-,
$P$- and $D$-partial waves in the \kpi\ channel. (Here both
longitudinally and transversely polarized $\psi^{\prime}$'s are
considered, and no attempt is made to restrict the strength of each
term to that seen for the $S$-, $P$- and $D$-wave \kpi\ components in
the data.) These curves show that although a peak can be made at
$\cos\theta_{\pi^-}\simeq 0.25$, it is necessarily accompanied by
large enhancements near $\cos\theta_{\pi^-}\simeq\pm 1$. No such
structures are evident in the \psitwospi\ mass plot of
Fig.~\ref{fig:z4430_mpipsip}.

\section{The \babar\ search for the \z\/}
\begin{figure}[htb]
\centering \includegraphics*[width=80mm]{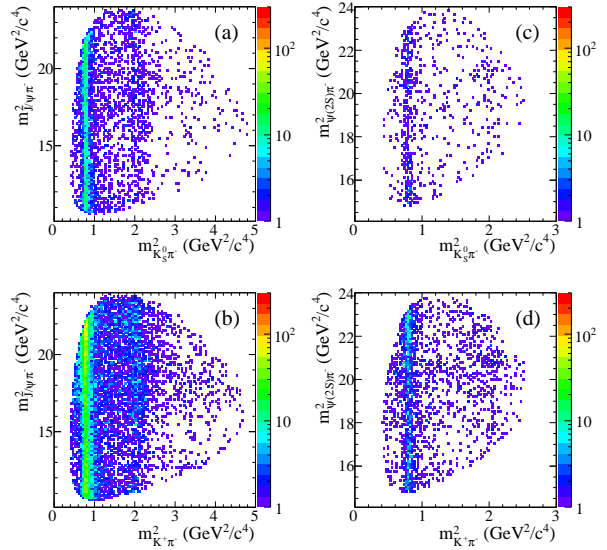}
\caption{From \babar\/~\cite{babar_z4430}, the $m_{\psi\pi^-}^2$
versus $m_{K\pi^-}^2$ Dalitz plots for the final samples in the decay
modes (a) $B^-\to J/\psi\pi^-K^0_S$, (b) $B^0\to J/\psi\pi^-K^+$, (c)
$B^-\to \psi^{\prime}\pi^-K^0_S$, and (d) $B^0\to
\psi^{\prime}\pi^-K^+$.}
\label{fig:Z4430_dp2}
\end{figure}

\begin{figure}[htb]
\centering \includegraphics*[width=80mm]{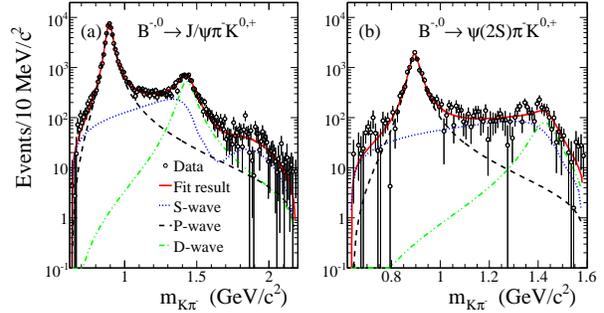}
\caption{From \babar\/~\cite{babar_z4430}, the fit results to the
\kpi\ mass distributions for the combined \kpi\ charge configurations
(a) $B^{-,0}\rightarrow J/\psi\pi^- K^{0,+}$ and (b)
$B^{-,0}\rightarrow \psi^{\prime}\pi^- K^{0,+}$. The open dots show the
data, and each curve represents different fit contribution, as
indicated.}
\label{fig:Jkpi_combined}
\end{figure}

\babar\ searched~\cite{babar_z4430} for the \z\ in the decay modes
$B\rightarrow\psi\pi^-K$, where $\psi=\jpsi$ or $\psi^{\prime}$ and
$K=K^+$ or $K^0_S$. The search was performed using the full \babar\
data set equivalent to integrated luminosity of 413 fb$^{-1}$
(455$\times 10^6$ $B\bar{B}$ candidates) collected at a center-of-mass
energy of 10.58 \gev\/. The \babar\ search covered two additional
decay modes ($B\rightarrow J/\psi\pi^-K$) that were not considered in
the Belle study. The $J/\psi$ modes contain about six times (see Table
II in Ref.~\cite{babar_z4430}) more data than those with the
$\psi^{\prime}$ and provide a powerful, high statistics tool for
understanding the background. The Dalitz plots for the four decay
modes are shown in Fig.~\ref{fig:Z4430_dp2}. The \kpi\ mass
distribution shows a clear \Ksone\ signal in all decay modes, and a
visible \Kstwo\ signal in the modes with a $J/\psi$ but less
significant in the modes with $\psi^{\prime}$. Since the \kpi\ system
dominates the Dalitz plot, the \babar\ Collaboration studied the \kpi\
mass distribution and the angular dependence. Similar contributions
from the different wave intensities are obtained in the modes with
$K^0_S$ and $K^+$ (see Table III in Ref.~\cite{babar_z4430}) and,
therefore, the modes with $K^0_S$ and $K^+$ are combined. The \kpi\
mass distributions are fitted with the $S$-, $P$-, and $D$-wave
intensities as shown in Fig.~\ref{fig:Jkpi_combined}. Good fits to the
data are obtained. The study of the angular dependence shows a clear
asymmetry in $\cos\theta_K$, where $\theta_K$ is the angle between the
Kaon and $\psi$ meson in the \kpi\ rest frame (see Figs.~12~and~13 in
Ref.~\cite{babar_z4430}).

The \babar\ mass resolution at the \z\ mass is $7~(4)$ \mevcc\ for the
modes with $J/\psi$ ($\psi^{\prime}$) (see Fig.~6 in
Ref.~\cite{babar_z4430}). Each event is efficiency corrected, and the
conclusions are based on distributions obtained after background
subtraction. The average efficiency varies slightly as a function of
the \kpi\ mass but can differ significantly between the different
$\psi$ decay modes (see Fig.~7 in Ref.~\cite{babar_z4430}), however
there are significant variations in efficiency with $\cos\theta_K$ for
certain regions of \kpi\ mass (see Fig.~36 in
Ref.~\cite{babar_z4430}).

\begin{figure}[htb]
\centering \includegraphics*[width=80mm]{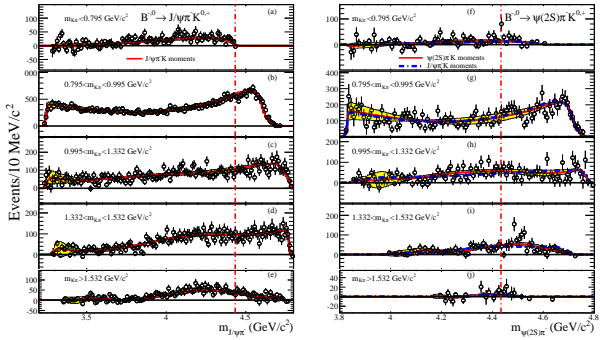}
\caption{From \babar\/~\cite{babar_z4430}, the \psipi\ mass
distributions in different regions of \kpi\ mass for (a-e)
$B^{-,0}\rightarrow J/\psi\pi^- K^{0,+}$, and (f-j)
$B^{-,0}\rightarrow \psi^{\prime}\pi^- K^{0,+}$; the open dots represent
the data, and the solid curves and shaded bands are due to the \kpi\
reflections in the different \psipi\ mass intervals, using the same
overall normalization constant. In (f-j), the dot-dashed curves are
obtained using \kpi\ normalized moments for $B^{-,0}\rightarrow
J/\psi\pi^- K^{0,+}$, instead of those from $B^{-,0}\rightarrow
\psi^{\prime}\pi^- K^{0,+}$; the vertical lines indicate
$m_{\psi\pi^-}=4.433$ \gevcc\/.}
\label{fig:ranges}
\end{figure}

\begin{figure}[!htbp]
\begin{center}
\includegraphics[width=80mm,height=60mm]{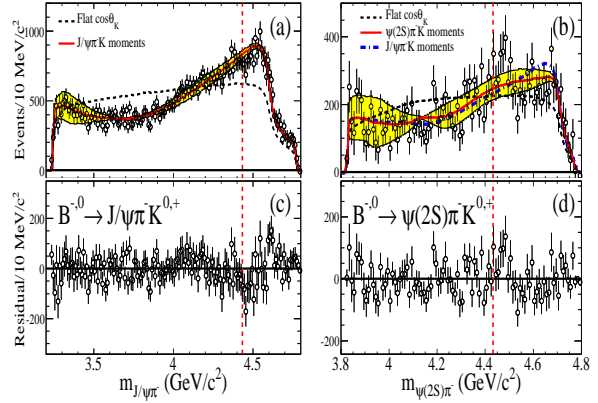}
\caption{From \babar\/~\cite{babar_z4430}, the \psipi\ mass
distributions for the combined decay modes (a) $B^{-,0}\rightarrow
J/\psi\pi^- K^{0,+}$ and (b) $B^{-,0}\rightarrow \psi^{\prime}\pi^-
K^{0,+}$. The points show the data (integrated over all \kpi\/
regions) after efficiency correction and background subtraction. The
dashed curves show the \kpi\ reflection for a flat \costhk\
distribution, while the solid curves show the result of \costhk\ when
accounting for the angular dependence (see Sec.~9 in
Ref.~\cite{babar_z4430}). The shaded bands represent the effect of
statistical uncertainty on the normalized moments. In (b), the
dot-dashed curve indicates the effect of weighting with the normalized
$J/\psi\pi^-K$ moments. The dashed vertical lines indicate the value
of $m_{\psi\pi^-}=4.433$ \gevcc\/. In (c) and (d), we show the
residuals (data-solid curve) for (a) and (b), respectively.}
\label{babarResults}
\end{center}
\end{figure}

The comparison between the \psipi\ mass distribution and the relevant
\kpi\ reflection was performed (see section 9 in
Ref.~\cite{babar_z4430}). This comparison is essential to investigate
the need for a \z\ signal above the \kpi\ reflection.  Following the
five \kpi\ regions (see section 10.D in Ref.~\cite{babar_z4430})
defined in the Belle analysis~\cite{belle_z4430}, the projection is
implemented in each \kpi\ interval as shown in
Fig.~\ref{fig:ranges}. In each figure, a comparison (not a fit)
between the data and the \kpi\ reflection is shown. Note that only one
normalization factor for the \kpi\ regions is used for each $\psi$
mode. The \kpi\ reflections onto the \psipi\ mass distributions
describe well the data, with a small statistical fluctuation at
$m_{K\pi^-}<0.795$ \gevcc\/. The $\psi\pi^-$ mass distribution and the
overall \kpi\ projection are shown in Fig.~\ref{babarResults}. No
significant evidence for \z\/ has been obtained. The \kpi\ projection
onto the \psipi\ mass distribution is used as a non-peaking
background. With such a background shape and a BW function for a \z\
signal, a signal is obtained for the overall \kpi\ region, but with
shifted mass value. A similar small signal is obtained in the \Ksone\
and \Kstwo\ combined regions. A $\sim 2\sigma$ peak is obtained when
the \Ksone\ and \Kstwo\ regions are vetoed. No (or a negative) signal
for $Z(4430)^-\rightarrow J/\psi\pi^-$ was found.

\begin{figure}[!htbpp]
\begin{center}
\includegraphics[width=80mm]{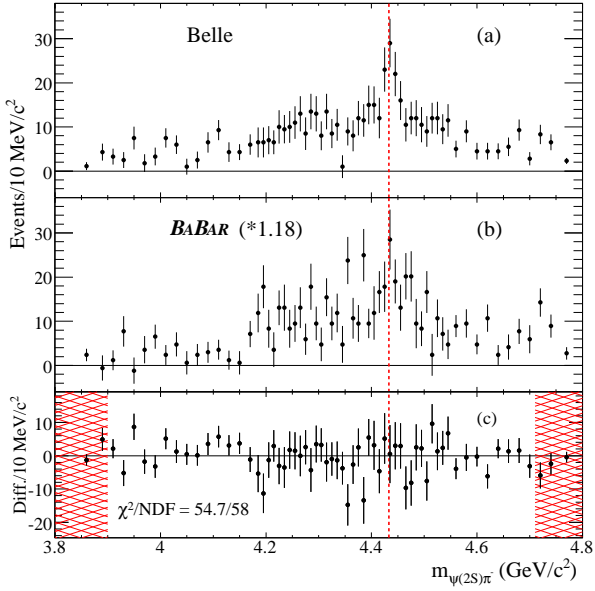}
\caption{From \babar\/~\cite{babar_z4430}, (a) The Belle \mpsitwospi\
distribution after background subtraction; (b) the equivalent \babar\
\mpsitwospi\ distribution obtained after scaling due to different
luminosities. (c) The difference between the \mpsitwospi\
distributions. The crosshatched area represents the exclusion region
for the normalization procedure. The vertical-dashed line indicates
$m_{\psip\pi^-}=4430$ \mevcc\/.}
\label{fig:diff2}
\end{center}
\end{figure}

Although the Belle Collaboration used a data sample equivalent to 1.46
times that of \babar\/'s, because of different selection criteria
\babar\ has more events per fb$^{-1}$. A direct comparison between the
two \mpsitwospi\ distributions is shown in Fig.~\ref{fig:diff2}. For
comparison purposes, the \babar\ distribution is not corrected for
efficiency, however both Belle and \babar\ distributions are
background subtracted and shown in Fig.~\ref{fig:diff2}(a) and
Fig.~\ref{fig:diff2}(b), respectively.  The \babar\ distribution is
scaled up by the factor 1.18 in order to normalize to the Belle data
between the crosshatched regions of Fig.~\ref{fig:diff2}(c). The
difference plot is shown in Fig.~\ref{fig:diff2}(c) where errors have
been combined in quadrature; here $\chi^2/NDF=54.7/58$, which
indicates that the two samples are statistically equivalent.

\section{Belle's Dalitz analysis of $B\to\psi^{\prime}\pi^-K$}
After the \babar\ Collaboration did not confirm the
$Z(4430)^-\rightarrow \psi^{\prime}\pi^-$ mass peak in their analysis
of $B\to \psi^{\prime}\pi^-K$ decays~\cite{babar_z4430}, the Belle
Collaboration performed a reanalysis~\cite{belle_z4430_dalitz} of
their data that took detailed account of possible reflections from the
\kpi\ channel. Specifically, they modeled the $B\to
\psi^{\prime}\pi^-K$ process as the sum of two-body decays $B\to
\psi^{\prime} K_i^{\ast}$, where $K_i^{\ast}$ denotes all of the known
$K^{\ast}\to$\kpi\ resonances that are kinematically accessible, and
both with and without a $B\to Z(4430)^- K$ component. The results of
this analysis, confirm the basic conclusions of Belle's initial
publication~\cite{belle_z4430}.

\begin{figure}[htb]
\centering \includegraphics*[width=80mm]{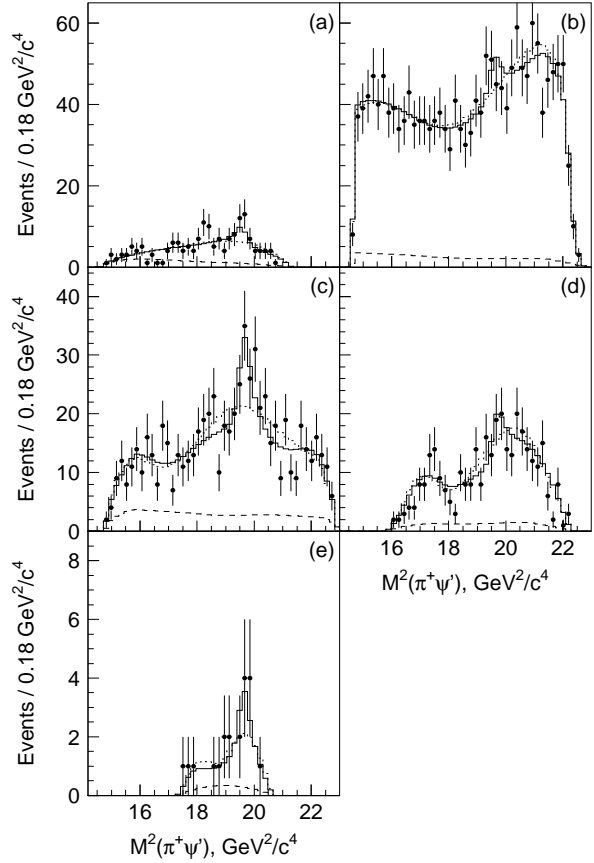}
\caption{From Belle~\cite{belle_z4430_dalitz},
$M^2_{\psi^{\prime}\pi^-} $ projections for $M_{K\pi}$ bands: (a)
below the \Ksone\/; (b) at the \Ksone\/; (c) between the \Ksone\ and
the $K^\ast_2(1430)$; (d) at the $K^\ast_2(1430)$; and (e) above the
$K^\ast_2(1430)$.  The dots show the data, the solid (dotted)
histograms are the results for the model with (without) a single
$\psi^{\prime}\pi^-$ state, and the dashed histograms are the
background contaminations.}
\label{fig:fig4abcde}
\end{figure}

\begin{figure}[htb]
\centering \includegraphics*[width=80mm]{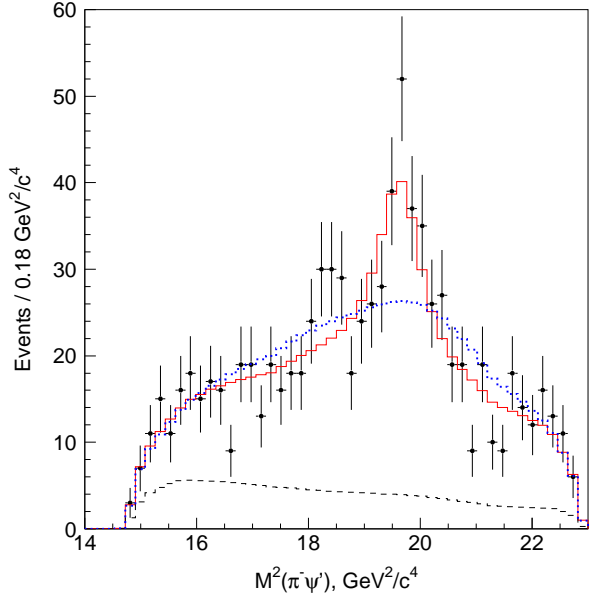}
\caption{From Belle~\cite{belle_z4430_dalitz}, the
$M^2_{\psi^{\prime}\pi^-}$ projection of the Dalitz plot with the
$K^\ast$ bands removed is shown as data points.  The histograms show
the corresponding projections of the fits with and without a
\z\/$\to\psi^{\prime}\pi^-$ resonance term.}
\label{fig:z4430_dalitz-analysis}
\end{figure}

Figure~\ref{fig:fig4abcde} shows the \psitwospi\ mass distribution in
different \kpi\ intervals as obtained from the Dalitz-plot analysis by
the Belle Collaboration~\cite{belle_z4430_dalitz}. In the region
between the \Ksone\ and the \Kstwo\ ($0.982<m_{K\pi}<1.332$ \gevcc\/,
Fig.~\ref{fig:fig4abcde}(c)), a \z\ peak is present. The data points
in Fig.~\ref{fig:z4430_dalitz-analysis} show
$M^2_{\psi^{\prime}\pi^-}$ Dalitz plot projections with the prominent
$K^\ast$ bands removed (as in Fig.~\ref{fig:z4430_mpipsip}) compared
with the results of the fit with no \z\ resonance, shown as a dashed
histogram, and that with a \z\ resonance, shown as the solid
histogram. The fit with the \z\ is favored over the fit with no \z\ by
$6.4\sigma$. The fitted mass, $M=4443^{+15~~+19}_{-12~~-13}$ \mevcc\/,
agrees within the systematic errors with the earlier Belle result; the
fitted width, $\Gamma = 107^{+86~~+74}_{-43~~-56}$ \mev\/, is larger,
but also within the analysis' systematic errors of the previous
result. In the default fit, the \z\ resonance was assumed to have zero
spin. Variations of the fit that included a $J=1$ assignment for the
\z\ as well as models that included additional, hypothetical $K^\ast
\to$\kpi\ resonances with floating masses and widths, and radically
different parameterizations of the \kpi\ $S$-wave amplitude do not
change the conclusions. The product branching fraction from the Dalitz
fit: ${\mathcal B}(B^0\to Z(4430)^- K)\times {\mathcal B}(Z(4430)^-
\to\psi^{\prime}\pi^-) = (3.2^{+1.8~+9.6}_{-0.9~-1.6})\times 10^{-5}$
is not in strong contradiction with the \babar\ 95$\%$ C.L. upper
limit of $3.1\times 10^{-5}$~\cite{babar_z4430}.

\section{Belle's two $Z$ peaks in the $\chi_{c1}\pi^-$ channel}
\begin{figure}[htb]
\centering
\includegraphics*[width=80mm]{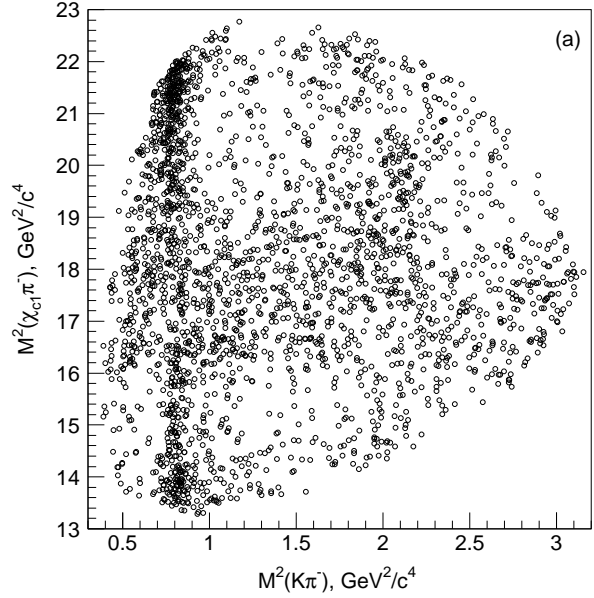}
\caption{From Belle~\cite{belle_z14050}, the $M^2_{K\pi^-}$
(horizontal) {\it vs.}  $M^2_{\chi_{c1}\pi^-}$ (vertical) Dalitz plot
distribution for candidate $B\to\chi_{c1}\pi^- K$ events.}
\label{fig:z4050_dalitz}
\end{figure}

In addition to the \z\/, Belle has presented results of an analysis of
$B\to\chi_{c1}\pi^-K$ decays that require two resonant states in the
$\chi_{c1}\pi^-$ channel~\cite{belle_z14050}. The $M^2_{K\pi^-}$ {\it
vs.} $M^2_{\chi_{c1}\pi^-}$ Dalitz plot, shown in
Fig.~\ref{fig:z4050_dalitz}, shows vertical bands of events
corresponding to \Ksone\/$\to$\kpi\ and \Kstwo\/$\to$\kpi\/, plus a
broad horizontal band near $M^2_{\chi_{c1}\pi^-}\simeq 17.5$
GeV$^2/\mathrm{c}^4$, indicating a possible resonance in the
$\chi_{c1}\pi^-$ channel. In this case, this horizontal band
corresponds to $\cos\theta_{\pi^-}\simeq 0$, a location where
interference between partial waves in the \kpi\ channel can produce a
peak and, thus, a detailed Dalitz analysis is essential.

For $B\to\chi_{c1}\pi^- K$, the kinematically allowed mass range for
the \kpi\ system extends beyond the $K^{\ast}_3(1780)$ $F$-wave
resonance and $S$-, $P$-, $D$- and $F$-wave terms for the \kpi\ system
are included in the model. The fit with a single resonance in the
$Z^-\to \chi_{c1}\pi^-$ channel is favored over a fit with only
$K^\ast$ resonances and no $Z^-$ by more than $10\sigma$. Moreover, a
fit with two resonances in the $\chi_{c1}\pi^-$ channel is favored
over the fit with only one $Z^-$ resonance by $5.7\sigma$. The fitted
masses and widths of these two resonances are: $M_1=4051\pm 14
^{+20}_{-41}$ \mevcc\ and $\Gamma_1 = 82^{+21~~+47}_{-17~~-22}$ \mev\
and $M_2=4248 ^{+44~+180}_{-29~~-35}$ \mevcc\ and $\Gamma_2 =
177^{+54~+316}_{-39~~-61}$ \mev\/. Here also, variations of the fit
that use a $J=1$ assignment for the $Z^-$ states and models that include
additional, hypothetical $K^\ast \to$\kpi\ resonances with floating
masses and widths, and different parameterizations of the \kpi\
$S$-wave amplitude do not change the conclusions.

\begin{figure}[htb]
\centering
\includegraphics*[width=80mm]{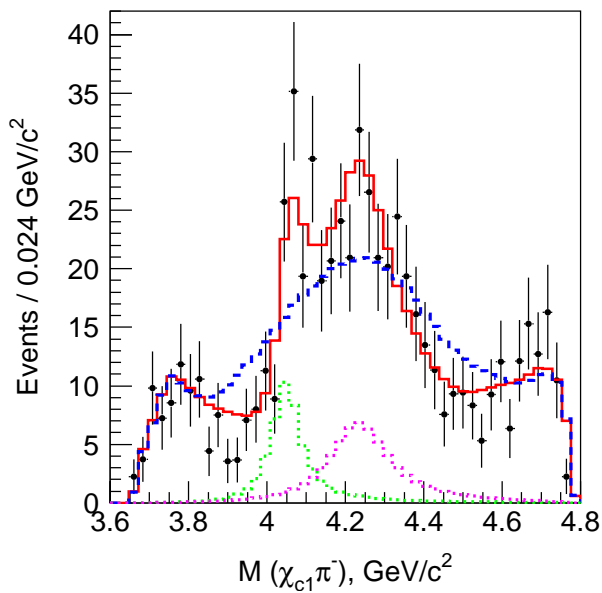}
\caption{From Belle~\cite{belle_z14050}, the data points show the
$M_{\chi_{c1}\pi^-}$ projection of the Dalitz plot with the $K^\ast$
bands removed. The histograms show the corresponding projections of
the fits with and without the two $Z\to\chi_{c1}\pi^-$ resonance
terms.}
\label{fig:z4050_dalitz-analysis}
\end{figure}

The product branching fractions for $B\to Z^-(\to\chi_{c1}\pi^-)K$
have central values similar to that for the \z\ but with large errors.
Figure~\ref{fig:z4050_dalitz-analysis} shows the $M_{\chi_{c1}\pi^-}$
projection of the Dalitz plot with the $K^\ast$ bands excluded and the
results of the fit with no $Z^-\to\chi_{c1}\pi^-$ resonances and with
two $Z^-\to\chi_{c1}\pi^-$ resonances.

\section{Summary}
If the peaks reported by Belle in the $\psi^{\prime}\pi^-$ and
$\chi_{c1}\pi^-$ channels are in fact meson resonances, they would be
``smoking guns'' for exotics. It is therefore important that the Belle
results are confirmed (or refuted) by other experiments. Both the
\babar\ and Belle \psitwospi\ analyses could benefit from additional
statistics. However, both experiments have completed data taking and
the \babar\ analysis used their complete dataset while the Belle
analysis used about 75$\%$ of their final total data sample.
Approximately 3 ab$^{-1}$ of integrated luminosity would be necessary
to bring the \psitwospi\ sample to the statistical level of \babar\/'s
current \jpsipi\ sample, and this amount of data will not be available
until the operation of one of the Super-$B$ factories, which will not
happen for at least a few years. In the meantime, it would be valuable
if \babar\ performed a search for the $Z_1(4050)^-$ and $Z_2(4250)^-$
in the $\chi_{c1}\pi^-$ channel and Belle reproduced \babar\/'s study
of the $\jpsi\pi^-$ channel using the existing data sets. The D0 and
CDF experiments at the Tevatron have data on hand that could be used
to carry out searches for inclusive production of the \z\ and we hope
that results will be available from them in the near future.

\section{Acknowledgment}
The work of Arafat Gabareen Mokhtar is supported by the US Department
of Energy, Division of High Energy Physics, Contract
DE-AC02-76-SF00515.  The work of Stephen Lars Olsen is supported by
the WCU program of the Korean Ministry of Education, Science and
Technology (grant number R32-2008-000-10155-0).


\begin{thebibliography}{99}
\bibitem{psiprime} The notations $\psi^{\prime}$ and $\psi(2S)$ are
used for the same charmonium state of $\psi(3686)$.
\bibitem{conj} The inclusion of charge-conjugate modes is always
implied.  Also, when two errors are presented, the first one is always
statistical and the second systematic.
\bibitem{theta-pi_def} $\theta_{\pi^-}$ is the angle between the
$\pi^-$ and the negative of the $\psi^{\prime}$ direction in the \kpi\ rest
frame.
\end{thebibliography}
\end{document}